\documentclass[twocolumn,aps,pra,showpacs,amsmath,amssymb,superscriptaddress]{revtex4-1}

\usepackage{physics}
\usepackage{graphicx} 
\usepackage[caption=false, labelformat=simple]{subfig}
\usepackage{dcolumn}
\usepackage{csquotes}
\usepackage[colorlinks=true,linkcolor = {blue}, citecolor = {blue}, urlcolor = {blue}]{hyperref}
\usepackage{xcolor}

\usepackage[export]{adjustbox}

\begin{document}

\title{Non-equilibrium dynamics and atom-pair coherence in strongly interacting Bose-Fermi mixtures}

\author{J. van de Kraats}
\email[Corresponding author:  ]{j.v.d.kraats@tue.nl}
\affiliation{Eindhoven University of Technology, P. O. Box 513, 5600 MB Eindhoven, The Netherlands}
\author{D.J.M. Ahmed-Braun}
\affiliation{TQC, Departement Fysica, Universiteit Antwerpen, Universiteitsplein 1, B-2610 Antwerpen, Belgium}
\author{V.E. Colussi}
\affiliation{Infleqtion, Inc., 3030 Sterling Circle, Boulder, CO 80301, USA}
\author{S.J.J.M.F. Kokkelmans}
\affiliation{Eindhoven University of Technology, P. O. Box 513, 5600 MB Eindhoven, The Netherlands}
\date{\today}

\begin{abstract}
Theoretical treatments of non-equilibrium dynamics in strongly interacting Bose-Fermi mixtures are complicated by the inherent non-Gaussian nature of the vacuum two-body physics, invalidating the typical Hartree-Fock-Bogoliubov approximation. Here, we apply the cumulant expansion to study non-equilibrium Bose-Fermi mixtures, which allows us to explicitly include the missing non-Gaussian quantum correlations, leading to a consistent dynamical theory of a Bose-Fermi mixture near an interspecies Feshbach resonance. We first apply our theory to a study of atom-pair coherence in the gas, which is significantly enhanced by the competing influences of the Fermi sea and Bose-Einstein condensation, in agreement with analytical calculations. Then, we study the depletion of a degenerate Bose-Fermi mixture following a quench to the unitary regime, characterizing the resulting depletion of the Bose-Einstein condensate, the deformation of the Fermi surface, and the production of molecules. We find that at early times, the population dynamics scale quadratically with the hold time, and define an associated characteristic timescale set by the parameters of the mixture and the width of the Feshbach resonance.
\end{abstract}

\maketitle

\section{INTRODUCTION}
\label{sec:intro}

A central problem in modern physics concerns the dynamics of many-body quantum systems far away from equilibrium. Especially challenging from a theoretical point of view is the description of such systems in the regime of strong interactions, associated with rapid generation of non-Gaussian correlations and a corresponding breakdown of the conventional quasiparticle approach \cite{Beliaev1958, Kaplan1976, Blazoit1985, Pitaevskii1997}. Understanding the underlying processes that drive this physics has implications for a wide array of fundamental questions, ranging from cosmology to nuclear and high-energy physics \cite{Kofman1997, Berges2018, Berges2021}.

In recent years, ultracold atomic gases near magnetic Feshbach resonances have emerged as a popular platform for probing non-equilibrium physics, both due to their excellent controllability in experiments, and their amicability to accurate theoretical modelling \cite{Chin2010, Kokkelmans2002, Kokkelmans2002_2, Gurarie2007}. Here, non-Gaussian correlations are essential for ergodicity \cite{Regemortel2018}, and can arise out of the presence of non-perturbative few-body physics, including novel three-body composites bound by the medium, and strongly enhanced three-body scattering by the presence of weakly bound trimer states due to the Efimov effect \cite{Levinsen2015, Fletcher2017, Klauss2017, Colussi2018, Musolino2019, Christianen2022, MusolinoPRL}. Recently, such effects were probed via a series of experiments exploring interaction quenches of degenerate Bose gases \cite{Makotyn2014, Eigen2017, Eigen2018}, where ergodicity is required for the system to transition from the degenerate to thermal regime. Inspired by these experiments, theoretical models were formulated based on a cumulant expansion of the interparticle correlations that develop following the quench \cite{Fricke1996, Kira2012, Kira2015, Colussi2020, Braun2022}, which were finally able to connect the relaxation dynamics in these systems to the prevalence of specific non-Gaussian correlations that encode three-body scattering physics \cite{Colussi2020, MusolinoPRL, Kraats2024}.

In light of this significant theoretical progress, it is natural to ask whether the same theoretical tools can be applied to study other systems where non-Gaussian correlations are known to play an important role. An interesting candidate, which is the subject of this work, is the Bose-Fermi mixture \cite{Viverit2000, Heiselberg2000, Yip2001, Viverit2002, Pu2002,  Stan2004, Inouye2004}. Historically, such systems were first studied in the context of \textsuperscript{3}He-\textsuperscript{4}He mixtures, which now find application in dilution refrigerators \cite{London1962, Hall1966, Zu2022}. In contemporary physics, Bose-Fermi mixtures have been used to study a wide array of physical phenomena, including anomalous pairing \cite{Avdeenkov2006, Watanabe2008, Fratini2010, Song2010, Song2011, Fratini2012}, phase separation \cite{Lous2018, Grochowski2020}, molecule formation \cite{Olsen2009, Cumby2013, Duda2023}, mediated interactions and sound propagation \cite{Kinnunen2015, DeSalvo2019, Edri2020, Manabe2021, Patel2023, Shen2024}, polarons \cite{Fritsche2021, Baroni2023}, and mixed superfluidity \cite{Ferrier2014, Roy2017}. 

In contrast to single-component Bose gases, where the vacuum two-body scattering physics is captured fully by Gaussian correlations, the equivalent physics in Bose-Fermi mixtures \emph{cannot} be obtained from a Gaussian approximation \cite{Bortolotti2006, Bortolotti2008}. This means that, while non-Gaussian physics in the single-component Bose gas arises predominantly from three-body effects, Bose-Fermi mixtures already show non-Gaussian signatures at the more fundamental two-body level. Physically, this distinction arises from the fact that the fermions in the gas do not condense, meaning that the interspecies correlation functions can not be reduced to a Gaussian form by the emergence of a classical condensate field. Similarly, the interaction between a BEC and a filled Fermi sea generates pairs at nonzero total momentum, described by non-Gaussian correlation functions. This contrasts with the unitary Bose gas, where pair excitations from the condensate create zero-momentum pairs, which are naturally Gaussian. Due to the associated theoretical difficulties, previous studies of the dynamics of Bose-Fermi mixtures rely on various simplifications, such as large population imbalances or reductions of the dimensionality of the system \cite{Wouters2003, Bortolotti2006, Olsen2009, Cumby2013, Garttner2019, Mistakidis2019, Abdullaev2019}.

Motivated by these outstanding theoretical challenges, and the recent progress in treating the non-Gaussian dynamics of the unitary Bose gas \cite{Colussi2020, MusolinoPRL, Kraats2024}, we develop in this work a cumulant model for the dynamics of a three-dimensional degenerate Bose-Fermi mixture near an interspecies Feshbach resonance. Our model respects all underlying conservation laws, is fully general with respect to the mass and density ratios of the mixture, and includes the effects of the finite width of the Feshbach resonance via a two-channel microscopic Hamiltonian widely used in treating resonant many-body systems \cite{Kokkelmans2002, Kokkelmans2002_2, Gurarie2007, Braun2022, Kraats2024}. Although our model can handle a variety of initial conditions, we benchmark it in this work by considering the simplest case where the gas starts out in the non-interacting ground state. Then, we study the correlation dynamics in the gas following a rapid quench of the interaction strength. Here we examine first the presence of atom-pair coherence in the post-quench dynamics, where fermions and bosons coherently oscillate between free and paired states. Next, we consider the depletion dynamics of the condensate following a quench to unitarity, where the interspecies scattering length diverges. Here we reveal how the spectrum of bosonic excitations that arises from the quench is uniquely shaped by Pauli blocking in the fermionic species. We also characterize the effect of the parameters of the mixture, such as the mass ratio and resonance width, which at early times can be expressed using a single characteristic timescale that sets the initial depletion of the condensate.

The paper is structured as follows. In Sec.~\ref{sec:model}, we develop our few- and many-body model, outlining the Hamiltonian, the calibration of the two-body interaction to the Feshbach resonance, and the cumulant equations of motion. Our results concerning atom-pair coherence and condensate depletion are presented in Secs.~\ref{sec:pairing} and \ref{sec:quench}. Finally, we conclude this paper in Sec.~\ref{sec:conc}.

\section{MODEL}
\label{sec:model}

\subsection{Hamiltonian}

We consider a uniform mixture of $N_B(N_F)$ identical bosons(fermions) of mass $m_B(m_F)$, coupled through an interspecies Feshbach resonance to diatomic molecules in a closed spin channel \cite{Chin2010}, with associated molecular particle number $N_M$. The total particle number follows as $N = N_B + N_F + 2N_M$, and we define total and species specific particle densities $n = N/V$ relative to the quantization volume $V$, such that $n = n_B + n_F + 2n_M$. For convenience we also introduce the relative two-body mass $m_R = m_B m_F/(m_B + m_F)$ and total two-body mass $M = m_B + m_F$. 

The essential microscopic physics of the system is captured by the Hamiltonian \cite{Kokkelmans2002, Gurarie2007, Bortolotti2006, Kraats2024},
\begin{eqnarray}
\hat{H} &=& \sum_{\vb{k}} \varepsilon_{\vb{k}}^{B} \hat{b}_{\vb{k}}^{\dagger} \hat{b}_{\vb{k}} + \sum_{\vb{k}} \varepsilon_{\vb{k}}^{F} \hat{f}_{\vb{k}}^{\dagger} \hat{f}_{\vb{k}} + \sum_{\vb{k}} \left(\varepsilon_{\vb{k}}^{M}  + \nu \right) \hat{m}_{\vb{k}}^{\dagger} \hat{m}_{\vb{k}} \nonumber \\ & +& \frac{v}{2V} \sum_{\vb{k}, \vb{k}', \vb{k}''} \zeta_{\vb{k}' - \vb{k} + 2\vb{k}''}\zeta_{\vb{k} - \vb{k}'} \ \hat{b}_{\vb{k}'+\vb{k}''}^{\dagger} \hat{b}_{\vb{k}-\vb{k}''}^{\dagger} \hat{b}_{\vb{k}}  \hat{b}_{\vb{k}'} \nonumber \\ & +&  \frac{g}{\sqrt{V}} \sum_{\vb{k}, \vb{k}'} \xi_{\frac{m_B}{M}\vb{k} + \vb{k}'}\left[\hat{m}_{\vb{k}}^{\dagger} \hat{b}_{-\vb{k}'} \hat{f}_{ \vb{k} + \vb{k}'} + \mathrm{h.c.} \right]. 
\label{eq:H}
\end{eqnarray}
Here operators $\hat{b}_{\vb{k}}$ and $\hat{f}_{\vb{k}}$ annihilate bosons and fermions at single-particle momenta $\vb{k}$, with associated kinetic energies $\varepsilon_{\vb{k}}^{B(F)} = k^2/2 m_{B(F)}$ ($\hbar = 1$). We have also defined the fermionic molecular operators \cite{Altman2005, Braun2022},
\begin{equation}
\hat{m}_{\vb{Q}} = \sum_{\vb{q}} \phi(q) \ \hat{b}_{\frac{m_R}{m_F}\vb{Q} + \vb{q}}' \hat{f}_{\frac{m_R}{m_B}\vb{Q} - \vb{q}}',
\end{equation}
where the primes denote annihilation of bosons and fermions in an energetically closed spin channel, and where $\phi(q)$ is the wave function of the closed-channel molecule as a function of the relative two-body momentum $q \equiv \abs{\vb{q}}$. The molecules have kinetic energy $\varepsilon_{\vb{Q}}^M = Q^2/2 M$, and bare internal energy $\nu$ relative to the scattering threshold. A Bose-Fermi pair can associate to form a molecule, controlled by a separable interaction potential with coupling strength $g$ and form factor $\xi_{\vb{q}} = \Theta(\Lambda_{BF} - \abs{\vb{q}})$. Here $\Theta(x)$ is the Heaviside step function such that $\Lambda_{BF}$ defines an ultraviolet cutoff on the relative momentum, interpreted physically as the inverse of the range of the interspecies interaction. 

From an exact solution of the two-body scattering problem, one finds that the parameters of the model are related to physical parameters of the Feshbach resonance as \cite{Kokkelmans2002, Braun2022, Kraats2024},
\begin{equation}
g^2 = \frac{\pi}{m_R^2 R_*}, \qquad \nu = \frac{1}{2 m_R R_*} \left(\frac{2 \Lambda_{BF}}{\pi} - \frac{1}{a_{BF}} \right).
\end{equation}
Here $R_*$ is the intrinsic length set by the molecular lifetime \cite{Petrov2004}, and $a_{BF}$ is the Bose-Fermi scattering length. Following Ref.~\cite{Ho2012}, we quantify the width of the Feshbach resonance using the dimensionless number $R_* k_F$, where $k_F = (6 \pi^2 n_F)^{1/3}$ is the Fermi momentum \cite{notekf}. Then, the limits $R_* k_F \ll 1$ and $R_* k_F \gg 1$ correspond to broad and narrow Feshbach resonances in the so-called many-body resonance width specification \cite{Ho2012}. In addition to the interspecies coupling, the bosons can also have a direct intraspecies interaction, which we model using a single-channel separable potential with coupling strength $v$ and a similar form factor $\zeta_{2\vb{q}} = \Theta(\Lambda_{BB} - \abs{\vb{q}})$. The coupling strength is defined by the Bose-Bose scattering length $a_{BB}$ as \cite{Kokkelmans2002},
\begin{equation}
\frac{1}{v} = \frac{m_B}{4\pi} \left(\frac{1}{a_{BB}} - \frac{2 \Lambda_{BB}}{\pi} \right).
\end{equation}
We consider only energy scales that are small enough such that the background $p$-wave interaction between fermions may be neglected \cite{Yip2001, Viverit2002}.

\subsection{Cumulant expansion and equations of motion}

\begin{table}[]
\begin{tabular}{ccp{40mm}}
\hline
\hline
Cumulant                                                                                                                     & Notation               & Interpretation                                                                           \\ \hline
$\expval*{\hat{b}_{\vb{0}}}_{\mathrm{c}}$                                                         & $\sqrt{V} \psi$         & Condensate wave function.                                                               \\
$\expval*{\hat{b}_{\vb{k}}^{\dagger} \hat{b}_{\vb{k}}}_{\mathrm{c}}$                                                         & $n_{\vb{k}}^B$         & Bosonic excitation  density.                                                                \\
$\expval*{\hat{f}_{\vb{k}}^{\dagger} \hat{f}_{\vb{k}}}_{\mathrm{c}}$                                                         & $n_{\vb{k}}^F$         & Fermionic density.                                                              \\
$\expval*{\hat{m}_{\vb{k}}^{\dagger} \hat{m}_{\vb{k}}}_{\mathrm{c}}$                                                         & $n_{\vb{k}}^M$         & Molecular density.                                                              \\
$\expval*{\hat{b}_{\vb{k}}\hat{b}_{-\vb{k}}}_{\mathrm{c}}$                                                                   & $\kappa_{\vb{k}}$    & Bosonic pairing.                                                                \\
$\expval*{\hat{f}_{\vb{k}}^{\dagger} \hat{m}_{\vb{k}}}_{\mathrm{c}}$                                                         & $\chi_{\vb{k}}$        & Atom-molecule conversion. \newline Closed-channel two-body\newline wave function. \\
$\expval*{\hat{b}_{\vb{k} - \vb{k}'}^{\dagger} \hat{f}_{\vb{k}'}^{\dagger} \hat{m}_{\vb{k}}}_{\mathrm{c}}$                     & $M_{\vb{k},\vb{k}'}$    & Atom-molecule conversion.                                                       \\
$\expval*{\hat{b}_{\vb{k} - \vb{k}'}^{\dagger} \hat{f}_{\vb{k}'}^{\dagger} \hat{f}_{\vb{k}}}_{\mathrm{c}}$ & $L_{\vb{k}, \vb{k}'}$   & Boson-fermion scattering. \newline Open-channel two-body \newline wave function.                                  \\ \hline \hline
\end{tabular}
\caption{\label{tab:cumulants} All cumulants included in our dynamical model, with a brief description of their physical interpretation. Note that by construction, a cumulant is only defined when all bosonic momenta are nonzero, except for the singlet $\expval*{\hat{b}_{\vb{0}}}_{\mathrm{c}}$.}
\end{table}

To track dynamics in the mixture we solve the Heisenberg equation of motion,
\begin{equation}
i \pdv{t} \expval*{\hat{O}} = \expval*{[\hat{O}, \hat{H}]},
\label{eq:HeisEOM}
\end{equation}
for expectation values of arbitrary many-body operators $\hat{O}$. We assume in general that the bosons form a Bose-Einstein condensate (BEC), which is modelled in the U(1) symmetry breaking picture by assuming that all bosons in the $\vb{k} = 0$ state form a coherent state such that $\expval*{\hat{b}_0} = \sqrt{V} \psi$, where $\psi$ is the wave function of the condensate. To track higher order correlations in the gas, we expand Eq.~\eqref{eq:HeisEOM} in a cumulant expansion \cite{Fricke1996,Kira2012, Kohler2002, Kokkelmans2002_2, Kohler2003, Kira2015}, which reduces expectation values to connected and irreducible few-body correlation functions. Previously, this method has been successfully applied to study the Efimov effect in the strongly interacting Bose gas, which manifests via a non-Gaussian three-body correlation function \cite{Colussi2020, MusolinoPRL, Kraats2024}. As simulating the full set of cumulants that desribe a given many-body system is numerically infeasible, a suitable truncation of the cumulant expansion is required. As higher order correlations develop sequentially following a quench from the non-interacting ground state, such a truncation will be valid provided that the model is limited to early times \cite{Kira2015, Colussi2020}, where the generation of excitations is limited. In addition, as we discuss further down below, including certain specific correlations may be required to ensure that the model remains physical in the sense that it obeys conservation of population and energy, or that it correctly reproduces few-body scattering processes in the vacuum or short-range limit. 

\begin{figure}
\includegraphics{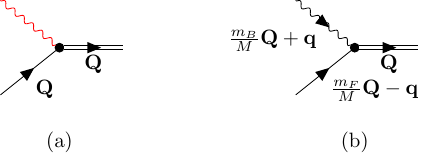}
\caption{\label{fig:KineticDiagrams} Two-body scattering diagrams that contribute to the density dynamics in the Bose-Fermi mixture, where red(black) wiggly lines represent (non)condensed bosons, and single(double) straight lines represent fermions(molecules). Process (a) represents scattering of fermions with the BEC, captured by $\chi_{\vb{Q}}$, while process (b) represents scattering of fermions with bosonic excitations, captured by $M_{\vb{Q},\frac{m_F}{M} \vb{Q} - \vb{q} }$. Here $\vb{Q}$ and $\vb{q}$ are the center-of-mass and relative momenta of the Bose-Fermi pair respectively.}
\end{figure}

The total set of cumulants that we include in our model is tabulated in Table~\ref{tab:cumulants}, including also their respective roles in the many-body dynamics of the mixture. This set contains in particular the excitation densities $n_{\vb{k}}$, normalized as,
\begin{eqnarray}
&&n_B = \abs*{\psi}^2 + \frac{1}{V} \sum_{\vb{k}} n_{\vb{k}}^B \equiv n_0^B + n_{\mathrm{exc}}^B, \\  && n_{F(M)} = \frac{1}{V} \sum_{\vb{k}} n_{\vb{k}}^{F(M)}.
\end{eqnarray}
Next to the densities themselves, we also need to include the atom-molecule conversion processes that will act to source the density dynamics. The lowest order such process, shown in Fig.~\ref{fig:KineticDiagrams}(a), describes the interaction between a fermion (solid black) and a boson in the BEC (wavy red), captured by the second order cumulant, or doublet, $\chi_{\vb{k}}$. To complete the set of doublet cumulants we define the bosonic pairing matrix $\kappa_{\vb{k}}$, which captures correlated pair excitations from the BEC. Note that this set excludes the fermionic and molecular pairing matrices $\expval*{\hat{f}_{\vb{k}} \hat{f}_{-\vb{k}}}_{\mathrm{c}}$ and $\expval*{\hat{m}_{\vb{k}} \hat{m}_{-\vb{k}}}_{\mathrm{c}}$, respectively, which always vanish provided the initial state for the fermionic component has a definite number of particles, and the Hamiltonian conserves the total fermionic population.

In identical Bose and two-component Fermi gases, a truncation of the cumulant expansion at the level of second order correlations, or doublets, produces the Hartree-Fock-Bogoliubov approximation \cite{Blazoit1985}, which is well studied in the literature and known to converge to the two-body Schr\"odinger equation in the short-range or vacuum limits \cite{Leggett2006}. In the Bose-Fermi mixture however, the same level of approximation leads to dynamical equations that do \textit{not} correctly reproduce the vacuum two-body physics, pointing out the fundamental shortcomings of the doublet approximation and the inherent non-Gaussian nature of the pairing physics in the Bose-Fermi mixture \cite{Bortolotti2006, Bortolotti2008}. As we will show formally in Sec.~\ref{sec:pairing}, this problem is remedied upon extending the doublet approximation with two non-Gaussian third order cumulants (triplets), defined in the last two rows of Table~\ref{tab:cumulants}. Here, $L_{\vb{k}, \vb{k}'}$ captures the triplet contribution to the cumulant expansion of the total two-body correlator $\expval*{\hat{b}^{\dagger} \hat{f}^{\dagger} \hat{b} \hat{f}}$, alternatively interpreted as the open-channel analogue of the doublet $\chi_{\vb{k}}$. In turn, $M_{\vb{k},\vb{k}'}$ captures the generalization of the scattering process in Fig.~\ref{fig:KineticDiagrams}(a) to the case where the boson is in an excited state, as shown in Fig.~\ref{fig:KineticDiagrams}(b). It is a necessary inclusion in the model as it ensures the conservation of the total system energy, or $\pdv{t}\expval*{\hat{H}} = 0$.

We now expand Eq.~\eqref{eq:HeisEOM} in the set of cumulants in Table~\ref{tab:cumulants}, which gives a set of coupled first-order differential equations that describe the dynamics of the system. For brevity we define the integrals,
\begin{equation}
\bar{n}_{\vb{k}}^B = \frac{1}{V} \sum_{\vb{k}'} \zeta_{\vb{k} - \vb{k}'}^2 n_{\vb{k}'}^B, \qquad \bar{\kappa} = \frac{1}{V} \sum_{\vb{k}} \zeta_{2\vb{k}} \kappa_{\vb{k}}.
\end{equation}

First, we have the extended Gross-Pitaevskii equation capturing the dynamics of the condensate wave function,
\begin{eqnarray}
i \dot{\psi} &=& v \left(\zeta_0^2 \abs*{\psi}^2 + 2 \bar{n}_{0}^B  \right) \psi + 
v \psi^* \zeta_0 \bar{\kappa} \nonumber \\ & +& \frac{g}{V} \sum_{\vb{k}} \xi_{\frac{m_B}{M} \vb{k}} \chi_{\vb{k}}.
\label{eq:GP}
\end{eqnarray}
Here the first line contains terms also found in the usual Gross-Pitaevskii equation for a single-component BEC with a contact potential of strength $v$, containing in addition the coupling to excited bosons and anomalous pairs through $n_{\vb{k}}^B$ and $\kappa_{\vb{k}}$ respectively \cite{Blazoit1985}. The second line couples the BEC to the fermionic component, through the scattering process shown in Fig.~\ref{fig:KineticDiagrams}(a). In turn, the dynamics of non-condensed bosons follows from,
\begin{eqnarray}
 \dot{n}_{\vb{k}}^B &=& 2 \mathfrak{I} \left( \Delta_{\vb{k}} \kappa_{\vb{k}}^{*} \right) \label{eq:nkB}\\ &+& 2 \frac{g}{\sqrt{V}} \sum_{\vb{k}'} \xi_{\vb{k} - \frac{m_B}{M}\vb{k}'} \mathfrak{I}\left(M_{\vb{k}', \vb{k}' - \vb{k}} \right) \nonumber \\
i  \dot{\kappa}_{\vb{k}} &=& 2 h_{\vb{k}}^{\mathrm{hf}} \kappa_{\vb{k}} + (1 + 2 n_{\vb{k}}^B) \Delta_{\vb{k}}.
\label{eq:kappaB}
\end{eqnarray}
Here we have defined the Hartree-Fock Hamiltonian and pairing field \cite{Blazoit1985}, quantifying the kinetic and interaction energies of bosonic pairs in the many-body background,
\begin{eqnarray}
h_{\vb{k}}^{\mathrm{hf}} &=& \varepsilon_{\vb{k}}^B + 2v \left( \zeta_{\vb{k}}^2 \abs*{\psi}^2 + \bar{n}_{\vb{k}}^B \right)  \\
\Delta_{\vb{k}} &=& v \zeta_{2\vb{k}} \left[ \zeta_{0} \psi^2 + \bar{\kappa} \right].
\end{eqnarray}
Note here that in the limit of vanishing Bose-Bose scattering length, $a_{BB} = 0$, we have $h_{\vb{k}}^{\mathrm{hf}} = \varepsilon_{\vb{k}}^B$ and $\Delta_{\vb{k}} = 0$. The coupling to the fermionic component is facilitated by the second line in Eq.~\eqref{eq:nkB}, representing the conversion of an excited boson and a fermion into a molecule, as also shown diagramatically in Fig.~\ref{fig:KineticDiagrams}(b). The analogous equations for the fermionic and molecular densities read,
\begin{eqnarray}
\dot{n}_{\vb{k}}^{F} &=& 2 g \xi_{\frac{m_B}{M}\vb{k}} \  \mathfrak{I} \left( \psi^* \chi_{\vb{k}} \right) \label{eq:nkF}\\ &+& 2\ \frac{g}{\sqrt{V}} \sum_{\vb{k}'} \xi_{\vb{k} - \frac{m_F}{M}\vb{k}'} \mathfrak{I} \left(M_{ \vb{k}', \vb{k}} \right),  \nonumber \\
 \dot{n}_{\vb{k}}^{M} &=& 2  g \xi_{\frac{m_B}{M}\vb{k}} \mathfrak{I} \left(\psi \chi_{\vb{k}}^{*} \right) \label{eq:nkM} \\ &-& 2 \ \frac{g}{\sqrt{V}} \sum_{\vb{k}'} \xi_{\frac{m_F}{M}\vb{k} - \vb{k}'} \mathfrak{I}\left(M_{\vb{k}, \vb{k}'} \right). \nonumber 
\end{eqnarray}
In both these equations the first and second line can be respectively connected to the diagrams in Figs.~\ref{fig:KineticDiagrams}(a) and (b), capturing the interactions with the BEC and excited bosonic fractions. The cumulant $\chi_{\vb{k}}$ evolves as,
\begin{eqnarray}
i  \dot{\chi}_{\vb{k}} &=&  (\varepsilon_{\vb{k}}^M + \nu - \varepsilon_{\vb{k}}^F) \chi_{\vb{k}} + g \psi \xi_{\frac{m_B}{M}\vb{k}} (n_{\vb{k}}^F - n_{\vb{k}}^M) \label{eq:chi}  \\ &+& \frac{g}{\sqrt{V}} \sum_{\vb{k}'} \xi_{\frac{m_F}{M}\vb{k} - \vb{k}'} L_{\vb{k}, \vb{k}'}^*. \nonumber   
\end{eqnarray}
As we will show in Sec.~\ref{sec:pairing}, this equation can be most easily interpreted as the closed-channel component of the two-body Schr\"odinger equation for Bose-Fermi pairs in the gas, and is thus an essential inclusion in the model for obtaining the correct pairing physics.

\begin{widetext}
Finally, the equations of motion for the triplets read,
\begin{eqnarray}
i  \dot{M}_{\vb{k}, \vb{k}'} &=& (\varepsilon_{\vb{k}}^M + \nu  - \varepsilon_{\vb{k} - \vb{k}'}^B -\varepsilon_{\vb{k}'}^F) M_{\vb{k}, \vb{k}'} - 2 v \left( \zeta_{\vb{k}-\vb{k}'}^2 \abs*{\psi}^2 + \bar{n}_{\vb{k} - \vb{k}'}^B \right) M_{\vb{k}, \vb{k}'}  \label{eq:M} \\ &+& \frac{g}{\sqrt{V}} \xi_{\frac{m_F}{M} \vb{k} - \vb{k}'} \left[(1 - n_{\vb{k}}^M) n_{\vb{k}- \vb{k}'}^{B} n_{\vb{k}'}^F  - (1 + n_{\vb{k} - \vb{k}'}^B)(1 - n_{\vb{k}'}^F) n_{\vb{k}}^M \right] 
+ g \xi_{\frac{m_B}{M}\vb{k}} \psi L_{\vb{k}, \vb{k}'}, \nonumber \\
i  \dot{L}_{\vb{k}, \vb{k}'} &=& (\varepsilon_{\vb{k}}^F - \varepsilon_{\vb{k} - \vb{k}'}^B - \varepsilon_{\vb{k}'}^F) L_{\vb{k}, \vb{k}'}  - 2 v \left( \zeta_{\vb{k} - \vb{k}'}^2 \abs{\psi}^2 + \bar{n}_{\vb{k} - \vb{k}'}^B \right) L_{\vb{k}, \vb{k}'} - v \zeta_{2\vb{k} - 2\vb{k}'} \left( \zeta_{0} (\psi^*)^2 + \bar{\kappa} \right)   L_{\vb{k}', \vb{k}}^* \label{eq:L} \\ &-& \frac{g}{\sqrt{V}} \xi_{\frac{m_F}{M}\vb{k} - \vb{k}'} (1 + n_{\vb{k} - \vb{k}'}^B - n_{\vb{k}'}^F) \chi_{\vb{k}}^{*} + \frac{g}{\sqrt{V}} \xi_{\vb{k} - \frac{m_F}{M}\vb{k}'} \kappa_{\vb{k} - \vb{k}'}^{*} \chi_{\vb{k}'}   + g \xi_{\frac{m_B}{M}\vb{k}} \psi^* M_{\vb{k}, \vb{k}'}. \nonumber 
\end{eqnarray}
\end{widetext}
Here $M$ captures the atom-molecule conversion processes shown in Fig.~\ref{fig:KineticDiagrams}b, as can be seen from the first term on the second line which takes the form of a Boltzmann type scattering term consistent with this diagram. The enhancement factors $(1 + n_{\vb{k} - \vb{k}'}^B)$ and $(1 - n_{\vb{k}'}^F),(1 - n_{\vb{k}}^M)$ capture respectively the Bose enhancement and Pauli blocking in this process arising from the Bose and Fermi statistics \cite{Kira2015}. The scattering cumulant $L$ is essentially the open-channel analogue of $\chi$, and can thus be interpreted as the open-channel component of the two-body wave function for Bose-Fermi pairs, as we will show in Sec.~\ref{sec:pairing}.

Up to this point, we have kept our model essentially general with respect to the background Bose-Bose interaction, demanding only that it is sufficiently weak such that a treatment at the level of the Hartree-Fock-Bogoliubov approximation is valid \cite{Colussi2020}. In the rest of this work however, we will limit ourselves to the case where the background interaction vanishes, i.e. $a_{BB} = 0$, focusing purely on the physics induced by the resonant interspecies interaction. We make this convenient simplification as the Bose-Bose interaction has no impact on the discussion of Bose-Fermi physics in Sec.~\ref{sec:pairing}, and only imparts quantitative shifts on the depletion dynamics discussed Sec.~\ref{sec:quench} without significantly affecting the qualitative features on which we focus (see Appendix \ref{ap:aBB}). If however one is interested instead in the effects of the Fermi sea on properties of the BEC such as the sound velocity or structure factor, a Bose-Bose interaction can not be neglected \cite{Shen2024}, which motivates our choice to include it in Eqs.~\eqref{eq:GP} - \eqref{eq:L} for generality.

\section{EMBEDDED TWO-BODY PHYSICS}
\label{sec:pairing}

As noted in the previous section, inclusion of the triplet cumulants $L_{\vb{k}, \vb{k}'}$ and $M_{\vb{k}, \vb{k}'}$ is necessary to ensure a correct embedding of the resonant two-body physics in the mixture, where ``correct embedding" implies that the two-body Schr\"odinger equation is contained within the many-body equations of motion, and can be recovered upon taking appropriate limits. In this section, we study the physics of strongly interacting Bose-Fermi pairs interacting in a many-body background consisting of a BEC and a Fermi sea. We subsequently show how this physics emerges from our equations of motion, and how it can be recognized dynamically through the presence of atom-pair coherence following a quench to the strongly interacting regime. For clarity we will always follow the convention of Fig.~\ref{fig:KineticDiagrams}, denoting the two-body center-of-mass and relative momenta as $\vb{Q}$ and $\vb{q}$ respectively.

\subsection{Static pair energies}

Before studying dynamics, we give a brief overview of relevant time-independent solutions to the embedded two-body Schr\"odinger equation. We will build up the complexity of the solution step by step, starting from an isolated Bose-Fermi pair interacting in vacuum, and subsequently adding in the many-body background. As we will show, it will be useful here to distinguish between stationary pairs whose total momentum $\vb{Q} = 0$, and moving pairs with $\vb{Q} \ne 0$.

For an interacting Bose-Fermi pair in vacuum, the Hamiltonian [Eq.~\eqref{eq:H}] can be diagonalized analytically to obtain the binding energy $E_{\mathrm{vac}}$ of the dressed Feshbach dimer, resulting in the integral equation \cite{Kokkelmans2002},
\begin{eqnarray}
E_{\mathrm{vac}} - \nu = \frac{g^2}{V} \sum_{\vb{q}} \xi_{\vb{q}}^2 \frac{1}{E_{\mathrm{vac}} - \varepsilon_{\vb{q}}^R},
\label{eq:PairE_bare}
\end{eqnarray}
where $\varepsilon_{\vb{q}}^R \equiv \varepsilon_{\vb{q}}^B + \varepsilon_{\vb{q}}^F$ is the relative kinetic energy of the Bose-Fermi pair. For $a_{BF}^{-1} > 0$ Eq.~\eqref{eq:PairE_bare} has a real solution $E_{\mathrm{vac}} < 0$, which represents the binding energy of the Feshbach dimer and obeys the universal prediction $E_{\mathrm{vac}} = -1/2 m_R a_{BF}^2$ near resonance.

\begin{figure}
\includegraphics{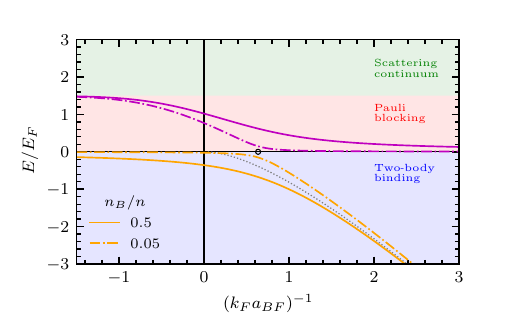}
\caption{\label{fig:PairE_Q0} Energy spectrum of a single Bose-Fermi pair with $\vb{Q} = 0$ embedded in a BEC and Fermi sea, as a function of the interspecies scattering length, with $R_* k_F = 0.5$ and $m_B = 2m_F$. Yellow and purple lines show the Feshbach dimer and anomalous dimer state respectively, plotted for two different values of the bosonic density $n_B$. The point of avoided crossing $(k_F a_{BF})^{-1} = 2/\pi$ is highlighted with a circle. Coloured shading distinguishes the different regions of energy, including the true two-body binding region $E<0$ in blue, the region of Pauli blockade $0<E<E_{\vb{0}}^{\mathrm{thr}}$ in red, and the true two-body scattering continuum $E>E_{\vb{0}}^{\mathrm{thr}}$ in green. Grey dotted line shows the energy $E_{\mathrm{vac}}$ of the Feshbach dimer state in vacuum.}
\end{figure}

We now seek to extend Eq.~\eqref{eq:PairE_bare} to a Bose-Fermi pair embedded in a BEC + Fermi sea mixture. In a many-body system, the Hamiltonian \eqref{eq:H} is not analytically solveable in general. As pointed out in Ref.~\cite{Dukelsky2011} however, the problem may be reduced to a solveable form by neglecting coupling terms between pairs that have nonzero center-of-mass momentum $\vb{Q} \neq 0$, equivalent to the replacement $\hat{m}_{\vb{Q}} \rightarrow \hat{m}_{0} \delta_{\vb{Q}, 0}$ in the Hamiltonian (see App.~\ref{ap:zeroQ}). The exact eigenstates of this reduced many-body Hamiltonian that may be generated from our initial condition contain only a \textit{single} interacting pair, formed from a boson in the BEC and a fermion at $\vb{k} = 0$. The energy spectrum of this pair is obtained from the integral equation,
\begin{eqnarray}
E - \nu = g^2 \xi_{\vb{0}}^2 \frac{n_0^B}{E} + \frac{g^2}{V} \sum_{\vb{q}} \xi_{\vb{q}}^2 \frac{1 - n_{\vb{q}}^F}{E - \varepsilon_{\vb{q}}^R},
\label{eq:PairE_Q0}
\end{eqnarray}
which reduces to Eq.~\eqref{eq:PairE_bare} in the vacuum limit $n_B \rightarrow 0$ and $n_F \rightarrow 0$, see App.~\ref{ap:zeroQ} for more detail. The solutions $E$ of Eq.~\eqref{eq:PairE_Q0} are illustrated in Fig.~\ref{fig:PairE_Q0}, plotted as a function of the dimensionless ratio $(k_F a_{BF})^{-1}$ which measures the scattering length with respect to the Fermi momentum. Comparing with Eq.~\eqref{eq:PairE_bare} we observe that the BEC and Fermi sea have competing effects on the energy of the pair. First, the Fermi sea blocks all momentum states $k < k_F$, thus raising the threshold for open-channel two-body scattering to $E_{\vb{0}}^{\mathrm{thr}} = k_F^2/2 m_R$, a mechanism also responsible for Cooper pairing in metals \cite{Song2010, Song2011}. The associated enhancement of the two-body interaction makes that the Feshbach dimer no longer disappears at resonance, but rather shifts above the vacuum threshold $E = 0$ at positive scattering length $(k_F a_{BF})^{-1} = 2/\pi$, as can be directly derived from Eq.~\eqref{eq:PairE_Q0} in the limit $n_0^B \rightarrow 0$. If the scattering length is tuned further to the attractive side of the resonance, the dimer moves up through the Fermi sphere towards the raised scattering threshold. Here we emphasize that this state constitutes an exact stationary state of the many-body Hamiltonian, meaning that the dimer state has infinite lifetime across the Feshbach resonance.

Counter to the effective repulsive interaction induced by the Fermi sea, the strong Bose-enhancement of the interaction between the fermion at $\vb{k} = 0$ and the BEC induces instead an effective attractive interaction. This leads to a \textit{splitting} of the pair energy into two levels, with an avoided crossing at $(k_F a_{BF})^{-1} = 2/\pi$. The lower branch, henceforth referred to as the Feshbach dimer, has negative energy for all scattering lengths. In turn, the higher branch, henceforth referred to as the anomalous dimer state \cite{Song2011}, is always embedded in the Fermi sphere and has positive energy.

\begin{figure}
\includegraphics{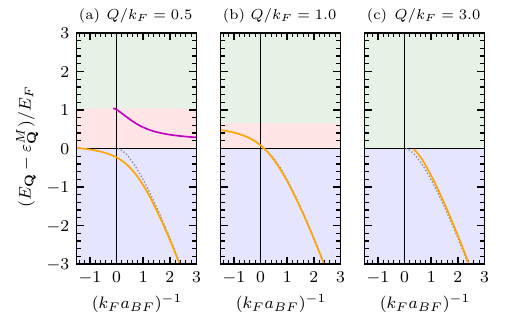}
\caption{\label{fig:PairE_Q} Energy spectra of embedded Bose-Fermi pairs as a function of interspecies scattering length, for different values of the center-of-mass momentum $\vb{Q}$. Layout of the plots is equivalent to Fig.~\ref{fig:PairE_Q0}. To focus on the binding energy we subtract the trivial center-of-mass energy $\varepsilon_{\vb{Q}}^M$. Computed with $R_* k_F = 0.5$, $m_B = 2 m_F$ and $n_B = n_F$, such that the Pauli blockade region vanishes completely at $Q/k_F = 3$, as shown in panel (c).}
\end{figure}

Eq.~\eqref{eq:PairE_Q0} can be generalised to the case $\vb{Q} \neq 0$ by deriving the many-body transition matrix in an extended ladder approximation that includes repeated scattering of fermions with the condensate \cite{Albus2002, Song2010, Song2011}, for details see App.~\ref{ap:LadApprox}. By solving for the poles of the transition matrix, one finds an integral equation for the embedded pair energy $E_{\vb{Q}}$,
\begin{eqnarray}
&& E_{\vb{Q}} - \varepsilon_{\vb{Q}}^M - \nu = g^2 \xi_{\frac{m_B}{M} \vb{Q}}^2 \frac{n_0^B}{E_{\vb{Q}}  - \varepsilon_{\vb{Q}}^F} \label{eq:PairE_Q} \\ &+& \frac{g^2}{V} \sum_{\vb{q}} \xi_{\vb{q}}^2   \frac{\Theta(\abs{\frac{m_F}{M} \vb{Q} - \vb{q}}-k_F)}{E_{\vb{Q}}  - \varepsilon_{\vb{Q}}^M - \varepsilon_{\vb{q}}^R} = 0 \nonumber,
\end{eqnarray}
which reduces to Eq.~\eqref{eq:PairE_Q0} in the limit $\vb{Q} \rightarrow 0$. The solutions of this equation for different values of $\vb{Q}$ are illustrated in Fig.~\ref{fig:PairE_Q}, for a particular choice of the mass ratio. Note here that, in contrast to the $\vb{Q} = 0$ reduction, the spectrum is now derived from an \textit{approximate} many-body transition matrix, in which the pair energies are real valued and the corresponding states still have infinite lifetime. In the full system however, these pair states should attain a finite lifetime due to the appearance of decay channels via scattering with particles in the many-body medium. For further discussion of the stability of the pair states, also in the scattering continuum, we refer to Ref.~\cite{Avdeenkov2006}.

In Fig.~\ref{fig:PairE_Q} we observe that a nonzero $\vb{Q}$ shifts the region of Pauli blocking and thus alters the two-body scattering threshold to \cite{Song2011},
\begin{eqnarray}
E_{\vb{Q}}^{\mathrm{thr}} = \varepsilon_{\vb{Q}}^M + \varepsilon_{\vb{q}_m}^R, \quad q_m = \mathrm{max}\left(k_F - \frac{m_R}{m_B} Q, 0 \right).
\end{eqnarray}
Hence when $Q = m_B k_F/m_R$, the effects of Pauli exclusion are fully cancelled and we recover the vacuum two-body scattering threshold $E_{\vb{Q}}^{\mathrm{thr}} = \varepsilon_{\vb{Q}}^M$, as shown in Fig.~\ref{fig:PairE_Q}(c). In addition, the asymptotic energy of the anomalous dimer state for small positive scattering lengths is pushed up from $E = 0$ to $E = \varepsilon_{\vb{Q}}^F$. At a critical value $\vb{Q} = k_F$, shown in Fig.~\ref{fig:PairE_Q}(b), $\varepsilon_{\vb{Q}}^F = E_{\vb{Q}}^{\mathrm{thr}}$ meaning that the anomalous dimer is shifted into the continuum for all values of the scattering length, and only the Feshbach molecule remains. 

We reiterate that these results are obtained from an approximate form of the many-body transition matrix, and it is known that higher order processes such as interactions between the BEC and virtual particle-hole excitations near the Fermi surface, can further shift the energy of Bose-Fermi pairs \cite{Song2011}.

\subsection{Quasi-stationary approximation}
\label{sec:QStat}
Now that we have established the pairing physics within the mixture, we can examine whether this physics is correctly contained within the equations of motion formulated in Sec.~\ref{sec:model}. For that we will focus on the dynamics of the cumulants $L_{\vb{Q},\frac{m_F}{M}\vb{Q} - \vb{q}}^*$ and $\chi_{\vb{Q}}$, interpreted as open and closed-channel components of the embedded two-body wave function. This is most easily shown by taking Eqs.~\eqref{eq:chi} and \eqref{eq:L} and neglecting inhomogeneous source terms from the BEC and bosonic pairing matrix by setting $\psi = \kappa_{\vb{k}} = 0$ \cite{Kira2015}. We then obtain the coupled equations of motion,
\begin{eqnarray}
i  \dot{\chi}_{\vb{Q}} &=&  (\varepsilon_{\vb{Q}}^M + \nu - \varepsilon_{\vb{Q}}^F) \chi_{\vb{Q}} \label{eq:QS} \\ & +&  \frac{g}{\sqrt{V}} \sum_{\vb{q}} \xi_{\vb{q}} L_{\vb{Q},\frac{m_F}{M}\vb{Q} - \vb{q}}^*, \nonumber \\
i  \dot{L}_{\vb{Q},\frac{m_F}{M}\vb{Q} - \vb{q}}^* &=& (\varepsilon_{\vb{Q}}^M + \varepsilon_{\vb{q}}^R - \varepsilon_{\vb{Q}}^F) L_{\vb{Q},\frac{m_F}{M}\vb{Q} - \vb{q}}^*   \\ &+& \frac{g}{\sqrt{V}} \xi_{\vb{q}} (1 + n_{\frac{m_B}{M}\vb{Q} + \vb{q}}^B - n_{\frac{m_F}{M}\vb{Q} - \vb{q}}^F) \chi_{\vb{Q}}. \nonumber
\end{eqnarray}
Here we have assumed that the dynamics of the densities are much slower than the dynamics of $L_{\vb{Q},\frac{m_F}{M}\vb{Q} - \vb{q}}^*$ and $\chi_{\vb{Q}}$, such that they may be approximated as constant, also known as the quasi-stationary approximation \cite{Colussi2018}. To solve Eqs.~\eqref{eq:QS} we adopt the ansatz,\color{black}
\begin{equation}
\begin{bmatrix}
L_{\vb{Q},\frac{m_F}{M}\vb{Q} - \vb{q}}^* \\
\chi_{\vb{Q}}
\end{bmatrix} 
= e^{- i \Delta E_{\vb{Q}} t} 
\begin{bmatrix}
\Psi_{\vb{Q}, \vb{q}}\\
\Phi_{\vb{Q}}
\end{bmatrix},
\label{eq:QStat}
\end{equation}
where $\Delta E_{\vb{Q}} = E_{\vb{Q}} - \varepsilon_{\vb{Q}}^F$ quantifies the energy difference between both sides of the fundamental scattering diagram in Fig.~\ref{fig:KineticDiagrams}(a), renormalized such that the energy of the molecule equals the dressed dimer energy $E_{\vb{Q}}$. Eqs.~\eqref{eq:QS} can now be rewritten as eigenvalue equations,
\begin{eqnarray}
E_{\vb{Q}} \Psi_{\vb{Q}, \vb{q}} &=& (\varepsilon_{\vb{Q}}^M + \varepsilon_{\vb{q}}^R) \Psi_{\vb{Q}, \vb{q}} \label{eq:Psi_Qk}  \\ &+& \frac{g}{\sqrt{V}} \xi_{\vb{q}} (1 + n_{\frac{m_B}{M}\vb{Q} + \vb{q}}^B - n_{\frac{m_F}{M}\vb{Q} - \vb{q}}^F) \Phi_{\vb{Q}}, \nonumber \\
E_{\vb{Q}} \Phi_{\vb{Q}} &=& (\varepsilon_{\vb{Q}}^{M} + \nu) \Phi_{\vb{Q}} + \frac{g}{\sqrt{V}} \sum_{\vb{q}} \xi_{\vb{q}} \Psi_{\vb{Q}, \vb{q}}, \label{eq:Phi_Q}
\end{eqnarray}
for the pair energy $E_{\vb{Q}}$. Eqs.~\eqref{eq:Psi_Qk} and \eqref{eq:Phi_Q} are equivalent to the momentum-space two-body Schr\"odinger equation for the open- and closed-channel wave functions $\Psi_{\vb{Q},\vb{q}}$ and  $\Phi_{\vb{Q}}$ respectively, with the two-body interaction again enhanced by the factor $(1 + n_{\frac{m_B}{M}\vb{Q} + \vb{q}}^B - n_{\frac{m_F}{M}\vb{Q} - \vb{q}}^F)$. Note here that the single-particle momenta match the diagram in Fig.~\ref{fig:KineticDiagrams}b.

Eqs.~\eqref{eq:Psi_Qk} and \eqref{eq:Phi_Q} constitute a set of linear algebraic equations, which may be directly diagonalized to obtain the pair energies $E_{\vb{Q}}$. Assuming no bosonic excitations $n_{\vb{k}}^B = 0$ and an ideal Fermi sea $n_{\vb{k}}^F = \Theta(k_F - k)$, one then finds that $E_{\vb{Q}}$ obeys Eq.~\eqref{eq:PairE_Q}, provided that we also take the limit $n_0^B \rightarrow 0$.
This proves the correct embedding of the two-body Schr\"odinger equation for interactions between fermions and noncondensed bosons. The appropriate inclusion of the mean field effects due to the BEC originate from the source terms scaling with $\psi$, including an additional mutual coupling between $L_{\vb{Q},\frac{m_F}{M}\vb{Q} - \vb{q}}^*$ and $M_{\vb{Q},\frac{m_F}{M}\vb{Q} - \vb{q}}^*$. Due to the nonlinear nature of the associated differential equations however, these effects can only be confirmed numerically, as we will do in the next section.

\subsection{Atom-pair coherence}
\begin{figure}
\includegraphics{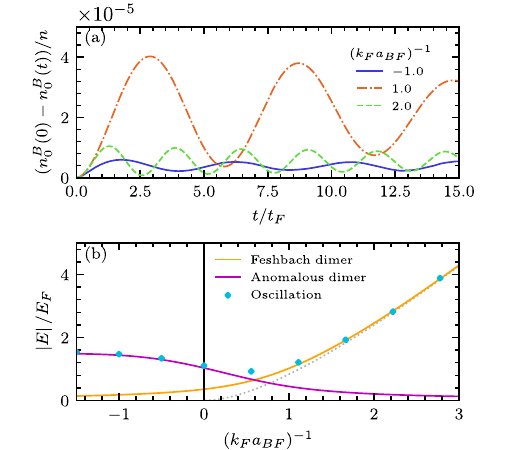}
\caption{\label{fig:CondOsc_PairE} Embedding of pair energies in the cumulant equations under the $\vb{Q} = 0$ approximation. In (a) we plot the depletion of the BEC as a function of time following an instantaneous quench to a given interspecies scattering length. In (b) we compare the frequencies of the resulting oscillation with the absolute energy of the Feshbach dimer and anomalous dimer as obtained from Eq.~\eqref{eq:PairE_Q0}. The vacuum dimer energy $\abs{E_{\mathrm{vac}}}$ is shown with a grey dotted line. Computed with mixture parameters $\Lambda_{BF}/k_F = 16.2$, $R_* k_F = 0.5$, $m_B = 2m_F$ and $n_B = n_F$.}
\end{figure}

To confirm the correct embedding of the pair energy in our equations of motion, we first adopt the $\vb{Q} = 0$ reduction of the Hamiltonian. As discussed in Sec.~\ref{sec:pairing}, this allows for the formation of just a single Bose-Fermi pair, thus protecting the Fermi sea and BEC from significant deformation/depletion. Furthermore, it will allow us to compare our numerical results directly with the exact solution of the many-body Hamiltonian as given in Eq.~\eqref{eq:PairE_Q0}. Starting from a pure BEC and Fermi sea, we instantaneously quench the system to a target scattering length $a_{BF}$, and subsequently track the evolution of the BEC population $n_0^B$. As shown in Fig.~\ref{fig:CondOsc_PairE}(a), we find that the population displays a coherent oscillation which we interpret as a Rabi-like exchange of atoms in the BEC with the $\vb{Q} = 0$ pair states plotted in Fig.~\ref{fig:PairE_Q0}. We note that similar oscillations have been observed experimentally through Rabi- and Ramsey-like magnetic field pulses, both in degenerate Bose gases \cite{Donley2002}, and non-degenerate Bose-Fermi mixtures \cite{Olsen2009}. To characterize the oscillations, we compute a discrete Fourier transformation of the condensate signal and subsequently extract the dominant oscillation frequency. Upon repeating this procedure for a set of target scattering lengths, we can compare directly with the pair energies computed in the previous section, see Fig.~\ref{fig:CondOsc_PairE}(b).

We observe that, for the full range of scattering lengths we study, the oscillations in the population of the condensate match well with the energies of the embedded pair, supporting our hypothesis that the oscillations seen in Fig.~\ref{fig:CondOsc_PairE}(a) originate from a coherent exchange of bosons between the BEC and the paired state. Especially interesting is that this coherence remains robust even for negative scattering lengths, where the exchange occurs between the BEC and the anomalous dimer in the Fermi sphere. This is a unique feature of the Bose-Fermi mixture, as a similar coherence in identical bose gases is not possible due to the absence of the Feshbach dimer for attractive interactions. At the crossing point $(k_F a_{BF})^{-1} = 2/\pi$, where the absolute energies of the Feshbach and anomalous dimer states are equal, we observe that the competing exchange between the two states leads to enhanced interference, resulting in a slight increase of the dominant oscillation frequency. For small negative scattering lengths, we find that the oscillations become increasingly incoherent, in accordance with the merging of the anomalous dimer with the two-body scattering continuum, to which the coherent oscillations show rapid dephasing as a function of time.

\begin{figure}
\includegraphics{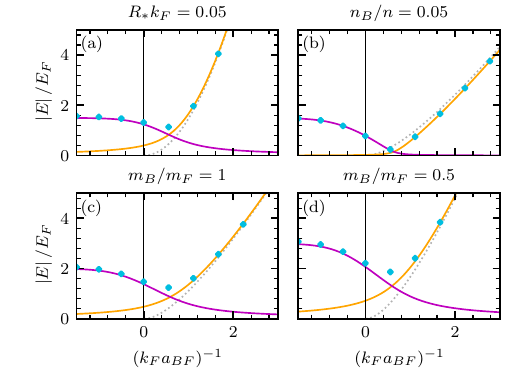}
\caption{\label{fig:CondOsc_PairE_arr} Comparison between oscillations in the BEC population and the pair energies in the $\vb{Q} = 0$ approximation, for different choices of the gas parameters. Layout and parameters match Fig.~\ref{fig:CondOsc_PairE}(b), except for a single changed parameter which is shown above the axes.}
\end{figure}
In Fig.~\ref{fig:CondOsc_PairE_arr}, we analyze the atom-pair coherence oscillations for different choices of the parameters of the mixture, again comparing between the analytical prediction in Eq.~\eqref{eq:PairE_Q0} and numerical simulations. In particular, we examine in Fig.~\ref{fig:CondOsc_PairE_arr}(a) a broader Feshbach resonance, leading to a steeper slope of the Feshbach dimer with scattering length. In Fig.~\ref{fig:CondOsc_PairE_arr}(b), we check that the mean field shift due to the BEC is accounted for correctly by decreasing the density of bosons, thus leading to a sharper avoided crossing consistent with Fig.~\ref{fig:PairE_Q0}(b). In Figs.~\ref{fig:CondOsc_PairE_arr}(c-d), we examine different mass ratios, which change both the slope of the Feshbach dimer and the two-body threshold energy $E_{\vb{0}}^{\mathrm{thr}}$. In all these cases, the match between our dynamical simulations and the analytical predictions remains robust, clearly showing the correct embedding of the two-body Schr\"odinger equation in our theory.

\begin{figure}
\includegraphics{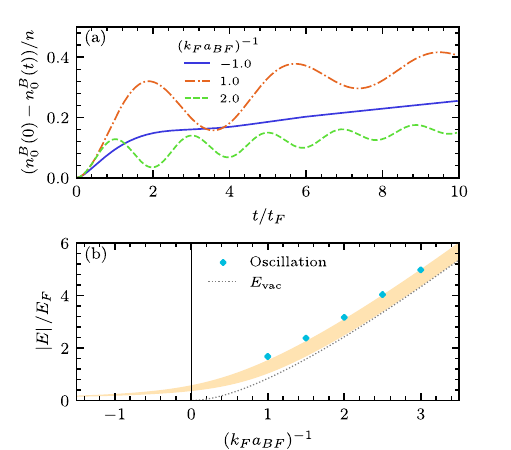}
\caption{\label{fig:CondOsc_PairE_Q} Atom-pair coherence in the Bose-Fermi mixture, including finite center-of-mass momentum pairs. In (a) we plot the depletion of the BEC as a function of time following an instantaneous quench to a set of interspecies scattering lengths. In (b) we compare the frequencies of the resulting oscillation with the absolute energy of the vacuum Feshbach dimer as obtained from Eq.~\eqref{eq:PairE_Q0}, and the range of pair energy differences $\Delta E_{\vb{Q}}$ for $Q \in \left[0, k_F \right]$ shown in yellow. Computed with mixture parameters $\Lambda_{BF}/k_F = 12.9$, $R_* k_F = 0.5$, $m_B = 2m_F$ and $n_B = n_F$.}
\end{figure}

The natural question to ask next is what happens to the atom-pair coherence once we relax the $\vb{Q} = 0$ restriction and use the full set of cumulant equations. It is intuitively expected that in this case the depletion of the BEC will no longer be microscopic, as the full Fermi sea is now allowed to interact with the condensate. For the same reason we should also expect that the majority of generated pairs have nonzero center-of-mass momentum, such that we start probing the $\vb{Q}$-dependent pair spectrum illustrated in Fig.~\ref{fig:PairE_Q}. With these points in mind, we plot in Fig.~\ref{fig:CondOsc_PairE_Q}(a) once more the depletion of the condensate as a function of time for a set of different scattering lengths. Interestingly, we find that while the depletion of the condensate is now macroscopic, for $a_{\mathrm{BF}} > 0$ it is still accompanied by a coherent oscillation that expresses a characteristic frequency, which we extract and plot as a function of scattering length in Fig.~\ref{fig:CondOsc_PairE_Q}(b). For $a_{\mathrm{BF}} <0 $ we no longer observe oscillations, which we attribute to the significant depletion of the Fermi sphere in this model (see Sec.~\ref{sec:quench}), thus negating the perfect Pauli blocking which leads to the appearance of the anomalous pair state in the $\vb{Q} = 0$ approximation.

The trend of the oscillations for $a_{\mathrm{BF}} > 0$ agrees well with the vacuum dimer energy, becoming higher in frequency as the scattering length is lowered. Quantitatively however, the oscillations of the condensate are faster than those of the vacuum dimer, which we attribute to an atom-pair coherence at nonzero center-of-mass momentum $\vb{Q}$. To extract the associated frequency, we consider the energy difference $\Delta E_{\vb{Q}}$ introduced for the quasi-stationary ansatz in Eq.~\eqref{eq:QStat}. From the initial condition $n_{\vb{k}}^F = \Theta(k_F - k)$ we expect that the majority of pairs are generated with $Q \in \left[0, k_F \right]$, so in Fig.~\ref{fig:CondOsc_PairE_Q} we plot the associated range of energies $\Delta E_{\vb{Q}}$, calculated directly from Eq.~\eqref{eq:PairE_Q}. We observe that for the majority of scattering lengths the oscillations in the BEC sit right at the edge of the range $\Delta E_{\vb{Q}}$, where $Q \approx k_F$, which we understand as a consequence of the large density of states for fermions near the fermi surface. For very large scattering lengths, near the resonance, we find that the oscillations shift outside of the expected range of $\Delta E_{\vb{Q}}$, which we attribute to the significant excitation of fermions above $k_F$ for these deep quenches, see our further quench results in Sec.~\ref{sec:quench}. We note that the rapid depletion of the Fermi sphere and BEC leads to a time-dependent oscillation frequency for these quenches, as the many-body background that embeds pairs is altered significantly through time. For all considered scattering lengths we additionally find that the interaction with the bosonic continuum through the scattering diagram in Fig.~\ref{fig:KineticDiagrams}(b) leads to a damping of the oscillation amplitude.

\section{DEPLETION DYNAMICS}
\label{sec:quench}

Having examined the embedded two-body physics, we now move on to study the density dynamics in our mixture following a deep quench to the strongly interacting regime, using the full set of cumulant equations derived in Sec.~\ref{sec:model}. Following earlier work in the unitary Bose gas, we keep our simulations limited to early times $t < 2 t_F$, as higher order correlations which we neglect in the current theory are expected to become non-negligible at longer times following the quench \cite{Colussi2020, Kraats2024}.

\subsection{Condensate depletion and molecule production}

\begin{figure}
\includegraphics{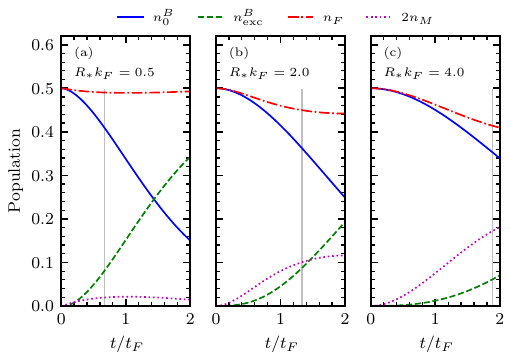}
\caption{\label{fig:Pops_Rskf} Dynamics of population fractions in the Bose-Fermi mixture following a deep quench of the non-interacting ground state to $(k_F a_{BF})^{-1} = 0$, for different values of the resonance strength $R_* k_F$. Vertical grey lines indicate the characteristic timescale $t_*/t_F$, as defined in the main text. Computed with mixture parameters $\Lambda_{BF}/k_F = 12.9$, $m_B = 2 m_F$ and $n_B = n_F$.}
\end{figure}
\begin{figure}
\includegraphics{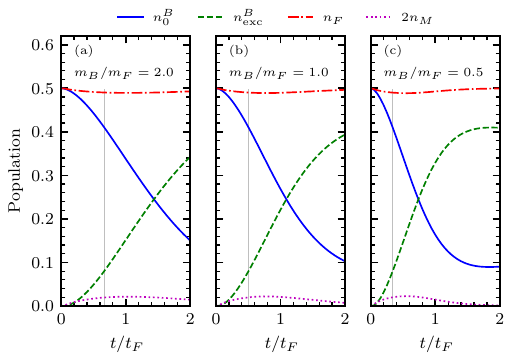}
\caption{\label{fig:Pops_mB} Dynamics of population fractions in the Bose-Fermi mixture following a deep quench of the non-interacting ground state to $(k_F a_{BF})^{-1} = 0$, for different values of the mass ratio $m_B/m_F$. Vertical grey lines indicate the characteristic timescale $t_*/t_F$, as defined in the main text. Computed with mixture parameters $\Lambda_{BF}/k_F = 12.9$, $R_* k_F = 0.5$ and $n_B = n_F$.}
\end{figure}

First, we examine in Figs.~\ref{fig:Pops_Rskf} and \ref{fig:Pops_mB} the depletion of the BEC and associated generation of bosonic excitations and fermionic molecules, which we find depends strongly on both the width of the Feshbach resonance and the mass ratio. We can quantify this observation more precisely by solving the cumulant equation of motion for $\chi_{\vb{k}}$ in the early time limit, capturing the scattering diagram in Fig.~\ref{fig:KineticDiagrams}(a). Equating the populations in Eq.~\eqref{eq:chi} with their initial ($t = 0$) values, $\psi = \sqrt{n_B}$, $n_{\vb{k}}^F = \Theta(k_F - k)$ and $n_{\vb{k}}^M = 0$, the resulting expression can be integrated exactly. Upon substituting the result into Eqs.~\eqref{eq:GP}, \eqref{eq:nkF} and \eqref{eq:nkM}, one finds
\begin{eqnarray}
n_0^B &=& n_B  - n_F f(t) , \label{eq:EarlyTime1}\\
n_{\vb{k}}^F &=& \Theta(k_F - k) \left[ 1 - f(t) \right], \nonumber \\
n_{\vb{k}}^M &=& \Theta(k_F - k) f(t), \nonumber
\end{eqnarray}
where we have defined a dimensionless time-dependent function,
\begin{eqnarray}
f(t) = \frac{4}{3 \pi}  \left(\frac{t}{t_*} \right)^2, \quad t_* = \sqrt{\frac{2 m_R}{m_F}} \sqrt{\frac{n_F}{n_B}} \sqrt{t_F \tau}
\label{eq:EarlyTime2}
\end{eqnarray}
Here we have followed Ref.~\cite{Braun2022} in defining a characteristic timescale $t_*$, dependent on the molecular lifetime $\tau = 2m_R R_*/ k_F$. By this definition, $t_*$ is interpreted as the density-averaged transition time for conversion of a Bose-Fermi pair in the open channel into a molecule in the closed channel. Since $\tau/t_F = R_* k_F m_R/m_F$, we see that for broad resonances ($R_* k_F \ll 1$) or small boson mass ($m_B \ll m_F$), $\tau \ll t_F$, such that molecules are rapidly generated following the quench. In the opposite limit of narrow Feshbach resonances ($R_* k_F \gg 1$) or large boson mass ($m_B \gg m_F$), $\tau \gg t_F$, the dynamics will instead be much slower.

% %
% \begin{figure}
% \includegraphics{mpop_resc.pdf}
% \caption{\label{fig:mpop_resc} Rescaled molecular population fraction $n_{M}^{\mathrm{resc}} \equiv n_M \cdot \left(t_*/t_*^{\mathrm{ref}} \right)^2$ following a quench to $(k_F a_{BF})^{-1} = 0$, for a set of resonance widths. Here $t_*^{\mathrm{ref}}$ is the value of $t_*$ for $R_* k_F = 2.0$. Remaining mixture parameters are $m_B = 2m_F$, $n_B = n_F$ and $v = 0$.}
% \end{figure}
% %
%
\begin{figure*}
\includegraphics{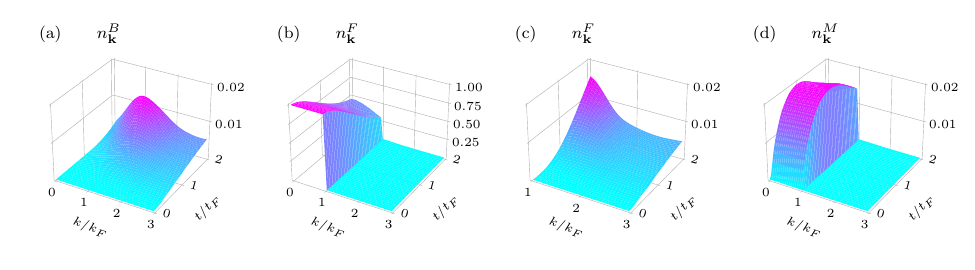}
\caption{\label{fig:nk_quench} Bosonic (a), fermionic (b,c) and molecular (d) momentum distributions following a quench to $(k_F a_{BF})^{-1} = 0$. Mixture parameters are $\Lambda_{BF}/k_F = 12.9$, $R_* k_F = 0.5$, $m_B = 2 m_F$ and $n_B = n_F$. In Fig. (c) we show a zoom of the fermionic distribution for $k>k_F$, which is too small to see on the scale of Fig. (b).}
\end{figure*}

While the scale $t_*$ dictates the transition time for processes of the type shown in Fig.~\ref{fig:KineticDiagrams}(a), Figs.~\ref{fig:Pops_Rskf} and \ref{fig:Pops_mB} show that, for broad resonances, the majority of bosons ejected from the BEC end up in excited states, whose dynamics are controlled by the cumulant $M_{\vb{k}, \vb{k}'}$ and the diagram in Fig.~\ref{fig:KineticDiagrams}(b). Although obtaining an analytic expression here is more complicated, our numerics show that the generation of bosonic excitations through this diagram occurs faster for broader resonances or smaller mass ratios $m_B/m_F$, once more consistent with the shorter lifetime of molecules. As our cumulant model assumes a macroscopically occupied BEC and a small excited state fraction, it follows that the temporal range of validity of the model, measured relative to the Fermi scale, increases for narrower Feshbach resonances or larger bosonic mass, where the atom-molecule transition rate $1/t_*$ becomes small. In the opposite limit, where the Feshbach resonance is broad or the bosonic mass is small, no such small parameter exists and the theory is limited to early times, similar to the case of the unitary Bose gas \cite{Braun2022, Kraats2024}. 

Our model predicts that at longer times the molecular population saturates to a maximum value, where the time to saturation scales with $R_* k_F$. Once saturation has occurred, the depletion of the BEC solely results in the generation of bosonic excitations. The saturated density of molecules generated in this time regime scales similarly with $R_* k_F$, which we interpret as a consequence of the increased closed-channel fraction for embedded dimers near narrow resonances. At even later times, $t \gg t_*$, we observe that the BEC population and bosonic excited state fraction also saturate, see Fig.~\ref{fig:Pops_mB}(c). In these regimes however we also expect higher-order correlations to significantly impact the dynamics, which could lead to further heating of the gas not captured by the current model \cite{Kraats2024}.

\subsection{Momentum distribution}

We now consider in more detail the dynamics of individual momentum modes following the quench. In Fig.~\ref{fig:nk_quench}, we plot the bosonic, fermionic and molecular momentum distributions through time. First, consider the molecular distribution $n_{\vb{k}}^M$, which we expect to be sourced directly following the quench via the scattering diagram in Fig.~\ref{fig:KineticDiagrams}(a), with the initial growth set by Eqs.~\eqref{eq:EarlyTime1} and \eqref{eq:EarlyTime2}. Consistent with these equations, we observe that $n_{\vb{k}}^M$ grows independently of momentum for all modes $k < k_F$, while the modes $k > k_F$ see negligible excitation. Consistent behavior can be recognized in the initial depletion of the Fermi sphere, as shown in Fig.~\ref{fig:nk_quench}(b). Hence, we see that following the quench a molecular Fermi surface develops, located at the momentum $k_F$ as set by the fermionic density.

At longer times following the quench, the dynamics set by the diagram in Fig.~\ref{fig:KineticDiagrams}(a) compete with the higher order scattering diagram in Fig.~\ref{fig:KineticDiagrams}(b), which generates bosonic excitations at $k > 0$ and fermionic excitations at $k> k_F$, see Figs.~\ref{fig:nk_quench}(a,c). Contrary to the unitary Bose gas, where $n_{\vb{k}}^B$ is typically a monotonously decreasing function \cite{Colussi2020}, the generation of bosonic excitations at momenta $k \ll k_F$ is suppressed by Pauli blockade from the Fermi sphere, which ensures that such low-momentum excitations can only be generated by high-momentum molecules with $Q \approx k_F$. In the bosonic distribution function, this effect competes with the universal decay of $n_{\vb{k}}^B$ at large momentum $k \gg k_F$ as set by the two-body contact \cite{Werner2012}, thus inducing a peak in $n_{\vb{k}}^B$ at $k \approx k_F$ where excitations can originate from all molecules inside the molecular Fermi sea. Hence, the same physics that leads to the appearance of an anomalous dimer state in the two-body spectrum discussed in Sec.~\ref{sec:pairing}, also imprints the presence of the Fermi sea on the bosonic excitation density following a quench, as can be clearly seen in Fig.~\ref{fig:nk_quench}(a).

\section{CONCLUSION}
\label{sec:conc}

In this work, we have introduced a novel model for the dynamics of Bose-Fermi mixtures near an interspecies Feshbach resonance, based on a cumulant expansion of the microscopic equations of motion. Our fully conserving theory goes beyond the typical Hartree-Fock-Bogoliubov approximation by including select non-Gaussian correlations, which we have shown are essential for capturing the physics of interacting pairs embedded in the many-body environment. In the Bose-Fermi mixture, such pairs exhibit a rich energy spectrum due to the competing effects of Bose enhancement from the BEC and Pauli blocking from the Fermi sphere, leading to the appearance of an anomalous dimer state in the vacuum two-body scattering continuum. Following a quench to a large interspecies scattering length, this spectrum directly determines the oscillation frequency of the populations in the gas, indicating coherent exchange of free atoms with bound molecules, or atom-pair coherence. These oscillations should be visible in a typical Ramsey pulse experiment \cite{Donley2002, Olsen2009}.

Subsequently, we have studied the population dynamics in a degenerate Bose-Fermi mixture following a deep quench of the interspecies interaction to the unitary regime, inspired by similar experiments with the unitary Bose gas \cite{Makotyn2014, Eigen2017, Eigen2018}. Here we find that the early time depletion of the condensate is characterized by a time scale $t_*$, interpreted as the density averaged atom-molecule transition time, which decreases for broader resonances or smaller boson mass. At longer times, we have studied how the persistent effects of the Pauli blockade due to the Fermi sphere affect the fundamental scattering diagrams in the gas, leading to a uniquely shaped bosonic excitation distribution with a peak around the Fermi momentum.

Looking ahead, our exploratory work suggests several interesting directions for future research, chief amongst these being the important question whether the deep quench dynamics that we have examined can also be seen in an experiment, where inite-size effects and non-coherent physics such as particle loss from the trap are also likely to affect observations. Furthermore, it is known from the unitary Bose gas that three-body correlations, enhanced by the Efimov effect, can strongly affect the heating dynamics of the gas for $t > t_F$ \cite{MusolinoPRL, Kraats2024}. As the Efimov effect also manifests in Bose-Fermi mixtures \cite{Sun2019}, it will be interesting to examine such correlations also in this system. Similarly, open questions remain regarding the effects of mediated Bose-Bose interactions in a dynamical setting \cite{DeSalvo2019, Edri2020, Shen2024}. Finally, our model may provide a new tool for studying the out-of-equilibrium dynamics of impurities, by taking the limit in which either the bosonic or fermionic density is very small, thus recovering the canonical problem of the Fermi or Bose polaron, which has recently seen significant experimental interest \cite{Fritsche2021, Skou2021, Skou2022, Baroni2023}.

\begin{acknowledgments}
We thank Leo Radzihovsky, Jacques Tempere, Michiel Wouters, Robert de Keijzer, Jasper Postema, Deon Janse van Rensburg, and Raul Santos for fruitful discussions. J.v.d.K. and S.J.J.M.F.K. acknowledge financial support from the Netherlands Organisation for Scientific research (NWO) under Grant No. 680.92.18.05, and from the Dutch Ministry of Economic Affairs and Climate Policy (EZK), as part of the Quantum Delta NL program. D.J.M.A.B acknowledges financial support by the
Research Foundation Flanders (FWO), Projects No. GOH1122N, No. G061820N, and No.
G060820N, and by the University Research Fund (BOF) of the University of Antwerp.
\end{acknowledgments}

\appendix

\section{EXACT SOLUTION FOR ZERO CENTER-OF-MASS MOMENTUM}
\label{ap:zeroQ}

As pointed out in Ref.~\cite{Dukelsky2011}, the Hamiltonian in Eq.~\eqref{eq:H} with $v=0$ becomes exactly solvable upon neglecting all coupling terms with nonzero center-of-mass momentum, or $\vb{Q} \neq 0$ in the notation of Sec.~\ref{sec:pairing}. Under this approximation, the Hamiltonian reduces to,
\begin{eqnarray}
\hat{H}_{\vb{Q} = 0} &=& \sum_{\vb{k}} \varepsilon_{\vb{k}}^{B} \hat{b}_{\vb{k}}^{\dagger} \hat{b}_{\vb{k}} + \sum_{\vb{k}} \varepsilon_{\vb{k}}^{F} \hat{f}_{\vb{k}}^{\dagger} \hat{f}_{\vb{k}} + \nu  \hat{m}_{\vb{0}}^{\dagger} \hat{m}_{\vb{0}} \\ & +& \frac{g}{\sqrt{V}} \sum_{\vb{k}} \xi_{\vb{k}} \left[\hat{m}_{\vb{0}}^{\dagger} \hat{b}_{\vb{k}} \hat{f}_{-\vb{k}} + \mathrm{h.c.} \right]. \nonumber
\end{eqnarray}
The exact many-body eigenstate $\ket*{\Psi}$, is given as,
\begin{eqnarray}
\ket*{\Psi} = \prod_{n = 1}^{N_p} \hat{\Gamma}_n^{\dagger} \ket*{\mathrm{UP}},
\end{eqnarray}
where $\ket*{\mathrm{UP}}$ is an \textit{unpaired} many-body Fock state, meaning that it contains no Bose-Fermi pairs with opposite single-particle momenta $(\vb{k}, - \vb{k})$, and is thus a trivial eigenstate of the Hamiltonian $\hat{H}_{\vb{Q} = 0}$. Additional eigenstates are constructed by the application of Bose-Fermi pair creation operators $\hat{\Gamma}_n^{\dagger}$, defined as,
\begin{eqnarray}
\hat{\Gamma}_{n}^{\dagger} = \sum_{\vb{q}} \xi_{\vb{q}} \frac{1}{\varepsilon_{\vb{q}}^r - E_n} \hat{f}_{\vb{q}}^{\dagger} \hat{b}_{-\vb{q}}^{\dagger} -\frac{\sqrt{V}}{g} \hat{m}_{\vb{0}}^{\dagger}. 
\end{eqnarray}
Here $\varepsilon_n$ is the pair energy, which obeys,
\begin{eqnarray}
E_n - \nu =  \frac{g^2}{V} \sum_{\vb{q}} \xi_{\vb{q}}^2 \frac{1 + \tilde{n}_{\vb{q}}^B - \tilde{n}_{\vb{q}}^F}{E_n - \varepsilon_{\vb{q}}^R},
\end{eqnarray}
where $\tilde{n}_{\vb{q}}^B, \tilde{n}_{\vb{q}}^F$ are the bosonic and fermionic momentum distributions in the unpaired state $\ket*{\mathrm{UP}}$. Note that each interacting pair has components in the open and closed spin channels, where the closed-channel component vanishes in the broad resonance limit $g \rightarrow \infty$. Assuming the initial condition of an ideal BEC and Fermi sea, as used in the main text, the only eigenstates accessible under unitary time evolution are those based on the unpaired state formed from the BEC plus a Fermi sea with a single hole in $\vb{k} = 0$, upon which is added a single interacting pair, with energy given by Eq.~\eqref{eq:PairE_Q0} in the main text.

\section{EXTENDED LADDER APPROXIMATION}
\label{ap:LadApprox}

\begin{figure}
\includegraphics[width = 0.46\textwidth]{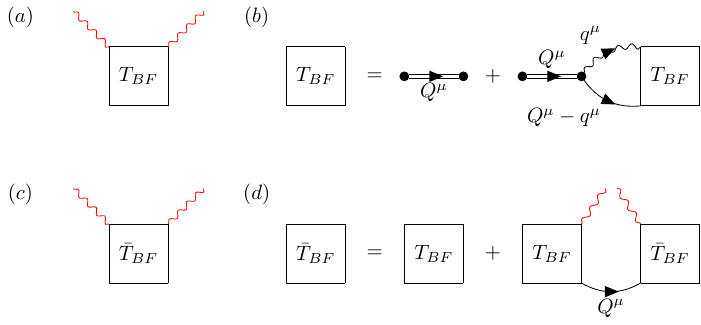}
\caption{\label{fig:LadApprox} Fermionic self-energy diagram (a) in the ladder approximation, with associated many-body transition matrix in (b). In the extended ladder approximation used in this work, the self energy diagram is extended to (c), with augmented many-body transition matrix given by (d). The (red) wavy lines denote (condensed) bosons, and the single(double) lines denote fermions(molecules). The vertices denote the Feshbach interaction, and we use four momenta $Q^{\mu} \equiv (\vb{Q}, \omega)$.}
\end{figure}

Here we derive the pair energy equation \eqref{eq:PairE_Q} for pairs with nonzero center-of-mass momentum, including the mean field shift induced by the BEC. First we formulate the many-body transition matrix $T_{BF}(\vb{q}_1, \vb{q}_2, \vb{Q}, \omega)$ at relative momenta $\vb{q}_1, \vb{q}_2$, total momentum $\vb{Q}$ and energy $\omega$, by using a standard ladder approximation for the Fermionic self energy $\Sigma_F$, as shown in Fig.~\ref{fig:LadApprox}. Solving the resulting Bethe-Salpeter equation \cite{Fetter2003} with our separable potential then gives the closed-form expression $T_{BF}(\vb{q}_{1}, \vb{k}_{q}, \vb{Q},\omega) =  \xi_{\vb{q}_1} \mathcal{T}_{BF}(\vb{Q}, \omega) \xi_{\vb{q}_2}$, where,
\begin{eqnarray}
 \mathcal{T}_{BF}(\vb{Q},\omega) =  \frac{1}{g^{-2} \left[G_M^0(\vb{Q}, \omega) \right]^{-1} - \Pi_{BF}(\vb{Q}, \omega)}.
\label{eq:LadApprox1}
\end{eqnarray}
Here,
\begin{eqnarray}
&&\Pi_{BF}(\vb{Q}, \omega) = \frac{i}{2 \pi V} \int_{-\infty}^{\infty} d\omega' \label{eq:LadApprox2} \\ && \qquad \sum_{\vb{k}} \xi_{\frac{m_B}{M} \vb{Q} - \vb{k}}^2  G_F^0(\vb{Q} - \vb{k}, \omega - \omega') G_B^0(\vb{k}, \omega'), \nonumber
\end{eqnarray}
and we have used the non-interacting Green's functions \cite{Fetter2003},
\begin{eqnarray}
G_F^0(\vb{k}, \omega) &=& \left[\frac{\Theta(q - k_F)}{\omega - \varepsilon_{\vb{k}}^F + i\eta} + \frac{\Theta(k_F - q)}{\omega - \varepsilon_{\vb{k}}^F-i\eta} \right], \\ G_B^0(\vb{k}, \omega) &=& \frac{1}{\omega - \varepsilon_{\vb{k}}^B + i\eta}, \nonumber \\   G_M^0(\vb{k}, \omega) &= &\frac{1}{\omega - \varepsilon_{\vb{k}}^M - \nu + i\eta}, \nonumber
\end{eqnarray}
where $\eta$ is a positive infinitesimal number. Eq.~\eqref{eq:LadApprox1} does not yet include the mean-field shift due to the BEC. To remedy this, we extend the ladder approximation of the fermionic self-energy to also include repeated scattering of the fermion with the BEC, see Fig.~\ref{fig:LadApprox}. Then we define an augmented transition matrix $\bar{T}_{BF}(\vb{q}_{1}, \vb{q}_{2}, \vb{Q},\omega) =  \xi_{\vb{q}_1} \bar{\mathcal{T}}_{BF}(\vb{Q}, \omega) \xi_{\vb{q}_2}$, whose associated integral equation shown in Fig.~\ref{fig:LadApprox} is solved to obtain,
\begin{equation}
\bar{\mathcal{T}}_{BF}(\vb{Q},\omega) =  \frac{1}{\mathcal{T}_{BF}^{-1}(\vb{Q}, \omega) - n_0^B \xi_{\frac{m_B}{M} \vb{Q}}^2 G_F^0(\vb{Q}, \omega)},
\end{equation}
Upon solving for the poles of $\bar{\mathcal{T}}_{BF}(\vb{Q},\omega)$, and performing the contour integration in Eq.~\eqref{eq:LadApprox2}, we obtain the pair energy equation \eqref{eq:PairE_Q} in the main text.

\section{NUMERICAL METHOD}
\label{ap:Numerics}

As a starting point, our numerical procedures are similar to the approach developed in Ref.~\cite{Colussi2020} for the unitary Bose gas. As the gas is uniform and the interaction isotropic, doublet cumulants are spherically symmetric and stored numerically as vectors on a discretized momentum grid. In turn, triplet cumulants depend on two absolute momenta, defined on the same grid, and the angle between them, which we similarly discretize. A triplet is then stored as a 3-dimensional matrix, and the right-hand side of the cumulant equations, as presented in Eqs. \eqref{eq:GP}-\eqref{eq:kappaB} and Eqs. \eqref{eq:nkF}-\eqref{eq:L}, can be evaluated using simple elementwise matrix operations. Sums over momenta are first transformed to the continuum limit relative to the quantization volume $V$, after which they are discretized by Simpson's 3/8 rule (cubic interpolation) \cite{Press2007}, thus accurately capturing the 3-dimensional volume element $\sim k^2$. 

To implement the cumulant equations, we also need to obtain permuted versions of the triplet cumulants. For example, the equation of motion for $n_{\vb{k}}^B$ \eqref{eq:nkB}, depends on an integral over the triplet $M_{\vb{k}', \vb{k}' - \vb{k}}$. However, the absolute momentum $\abs*{\vb{k}' - \vb{k}}$ and angle $\vu{k}' \cdot \widehat{\vb{k}' -\vb{k}}$ will typically not fall onto an exact grid point, requiring some interpolation method to obtain $M_{\vb{k}', \vb{k}' - \vb{k}}$ from the closest known elements of the cumulant $M_{\vb{k}, \vb{k}'}$. In Ref.~\cite{Colussi2020}, this problem was solved by nearest-neighbour interpolation, simply taking the value of $M_{\vb{k}, \vb{k}'}$ at the nearest grid point. This method is very fast since the 3D matrix $M_{\vb{k}', \vb{k}' - \vb{k}}$ can be completely obtained from a simple remapping of $M_{\vb{k}, \vb{k}'}$, without the need for any additional computation. In numerical testing however we have found the nearest-neighbour method to be insufficiently precise for accurately capturing the delicate balance between cumulants that ensures energy conservation in our model, which was not a concern in Ref.~\cite{Colussi2020} as energy conservation was already violated by the equations themselves. These issues are exacerbated by a difficulty that is unique to the Bose-Fermi model, namely the enormous discontinuity that exists in the initial condition due to the fermionic density $n_{\vb{k}}^F$, which is cut abruptly above the Fermi momentum. This essential feature of the system exacerbates any errors arising from the numerical discretization, meaning that more care is required.

To alleviate the problems above, we adjust the method of Ref.~\cite{Colussi2020} in a few ways. First, we ensure that triplet cumulants are defined in such a way that the discontinuity due to the Fermi sphere manifests in the momentum grid rather than the angular grids. For example, the cumulant $M_{\vb{k}, \vb{k}'}$ is defined such that $\vb{k}$ and $\vb{k}'$ are fermionic momenta, and thus see the discontinuity, while the third momentum in the cumulant, $\vb{k} - \vb{k}'$, is bosonic, meaning that the angular coordinate does not see the discontinuity. Hence the matrix $M_{\vb{k}, \vb{k}'}$ is discontinuous along $\abs*{\vb{k}}$ and $\abs*{\vb{k}'}$, but not along $\vu{k} \cdot \vu{k}'$. We then ensure that the momentum grid is defined such that the momentum $k_F$ falls exactly in the center of an integration interval set by Simpson's rule, such that the integral over the discontinuity would be exact if it were a perfect symmetric step function. We use a similar procedure for the angular grid in the permuted triplet $M_{\vb{k}', \vb{k}' - \vb{k}}$, which will be equally spaced around the angle $c_f$ where $\abs*{\vb{k}' - \vb{k}} = k_F$, if this angle exists. Finally, to further reduce discretization errors from permuting triplets, we compute $M_{\vb{k}', \vb{k}' - \vb{k}}$ by 3D linear interpolation instead of the nearest-point method. We adjust the interpolation to extrapolation from below(above) for points just below(above) the Fermi surface, thus ensuring that the discontinuity is not artificially smoothed by interpolation.

\section{NONZERO BOSE-BOSE INTERACTIONS}
\label{ap:aBB}
\begin{figure}
\includegraphics{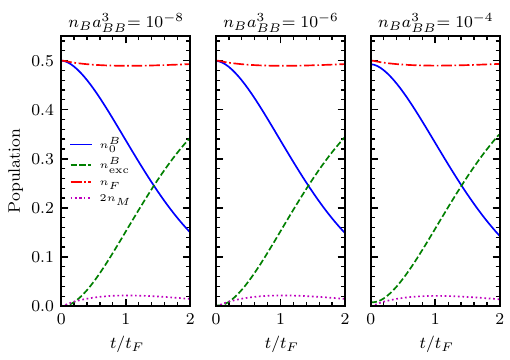}
\caption{\label{fig:nonzero_aBB} Dynamics of population fractions in the Bose-Fermi mixture following a deep quench to $(k_F a_{BF})^{-1} = 0$, for different values of the Bose-Bose scattering length. Computed with mixture parameters $\Lambda_{BF}/k_F = \Lambda_{BB}/k_F = 12.9$, $R_* k_n = 0.5$,  $m_B = 2 m_F$ and $n_B = n_F$.}
\end{figure}

As noted in Sec.~\ref{sec:model}, turning on a weak background interactions between the bosons does not qualitatively alter our results. We prove this assertion in Fig. \ref{fig:nonzero_aBB}, where we once more plot the population dynamics in the gas following a deep quench to unitarity, now with a slightly repulsive Bose-Bose interaction.
Here, we initialize the bosonic component of the mixture in the Bogoliubov ground state corresponding with the value of $a_{BB}$ \cite{Pitaevskii1997}, meaning that there is now a small fraction of excited bosons on top of the BEC. It is evident that a weak Bose-Bose interaction has negligible effect on the population dynamics discussed in the main paper, which can be straightforwardly understood from the dominance of the Bose-Fermi interaction near the Feshbach resonance. We note however, that $a_{BB}$ does have a significant effect on the bosonic pairing cumulant $\kappa_{\vb{k}}$, which may be relevant for more detailed studies of the effect of mediated Bose-Bose interactions on, for example, the dispersion of elementary excitations in the condensate. This is however beyond the scope of the present work, and we will leave it as an interesting opportunity for future study.

\bibliography{References}

\begin{thebibliography}{81}
\bibitem [{not()}]{notekf}%
  \BibitemOpen
  \href@noop {} {}\bibinfo {note} {Here we make the choice to use the fermion
  density to set a characteristic momentum scale, which will be a
  representative scale for the average {B}ose-{F}ermi interparticle separation
  provided that the boson and fermion densities are not too
  dissimilar.}\BibitemShut {Stop}%
\end{thebibliography}

\end{document}